\documentclass{appolb}
\usepackage{bm,graphicx}
\newcommand{\pvec}{\bm{p}}

\newcommand{\laph}{\textsc{Laph}}

\begin{document}
\title{Exploring Excited Hadrons in Lattice QCD
\thanks{Presented at Excited QCD, Zakopane, Poland, 8-14 February 2009}%
}
\author{Colin Morningstar\\(for the Hadron Spectrum Collaboration)
\address{Department of Physics, Carnegie Mellon University, 
        Pittsburgh, PA 15213, USA}
}
\maketitle
\begin{abstract}
Progress in extracting the spectrum of excited hadron resonances in
lattice QCD Monte Carlo calculations is reviewed
and the key issues and challenges in such computations are outlined.
The importance of multi-hadron states as simulations with lighter pion masses 
are done is discussed, and the need for all-to-all quark propagators is 
emphasized.
\end{abstract}
  
Experiments show that many excited-state hadrons exist, and there are significant
experimental efforts to map out the QCD resonance spectrum, such as Hall B
and the proposed Hall D at Jefferson Lab, ELSA associated with the University
of Bonn, COMPASS at CERN, PANDA at GSI, and BESIII in Beijing.
Hence, there is a great need for \textit{ab initio} determinations of such
states in lattice QCD.

Higher-lying excited hadrons are a new frontier in lattice QCD, and
explorations of new frontiers are usually fraught with dangers.  Excited states 
are more difficult to extract in Monte Carlo calculations; correlation matrices 
are needed and operators with very good overlaps onto the states of interest are
crucial.  To study a particular state of interest, all states lying below that
state must first be extracted, and as the pion gets lighter in lattice QCD 
simulations, more and more multi-hadron states will lie below the excited
resonances.  To reliably extract these multi-hadron states, multi-hadron 
operators made from constituent hadron operators with well-defined relative
momenta will most likely be needed, and the computation of temporal correlation
functions involving such operators will require the use of all-to-all quark 
propagators.  The evaluation of disconnected diagrams will ultimately be
required.  Perhaps most worrisome, most excited hadrons are unstable 
(resonances), so the results obtained for finite-box stationary-state
energies must be interpreted carefully.

Reliably capturing the masses of excited states requires the computation
of correlation matrices $C_{ij}(t)=\langle 0\vert T \Phi_i(t)
 \Phi^\dagger_j(0) \vert 0\rangle$ associated with a large set of $N$
different operators $\Phi_i(t)$.  It has been shown in Ref.~\cite{wolff90}
that the $N$ {\em principal effective masses} $W_\alpha(t)$, defined by
\[  W_\alpha(t)=\ln\left(\frac{\lambda_\alpha(t,t_0)}{
 \lambda_\alpha(t+1,t_0)}\right),
\]
where $\lambda_\alpha(t,t_0)$ are the eigenvalues of
 $C(t_0)^{-1/2}\ C(t)\ C(t_0)^{-1/2}$ and $t_0<t/2$ is usually chosen,
tend to the eigenenergies of the lowest $N$ states with which the
$N$ operators overlap as $t$ becomes large.  The eigenvectors
associated with $\lambda_\alpha(t,t_0)$ can be viewed as 
variationally optimized operators.  When combined
with appropriate fitting and analysis methods, such variational
techniques are a particularly powerful tool for investigating excitation
spectra.  

The use of operators whose correlation functions $C(t)$ attain their
asymptotic form as quickly as possible is crucial for reliably
extracting excited hadron masses.  An important ingredient in constructing
such hadron operators is the use of smeared fields.  Operators constructed
from smeared fields have dramatically reduced mixings with the high frequency
modes of the theory.  Both link-smearing and quark-field smearing should
be applied.  Since excited hadrons are expected to be large objects, 
the use of spatially extended operators is another key ingredient in
the operator design and implementation.  A more detailed discussion
of these issues can be found in Ref.~\cite{baryons1}.

Hadron states are identified by their momentum $\pvec$, intrinsic
spin $J$, projection $\lambda$ of this spin onto some axis,
parity $P=\pm 1$, and quark flavor content (isospin, strangeness, {\it etc.}).
Some mesons also include $G$-parity as an identifying quantum number.
If one is interested only in the masses of these states, one can restrict
attention to the $\pvec=\bm{0}$ sector, so operators must be invariant 
under all spatial translations allowed on a cubic lattice.  The little 
group of all symmetry transformations on a cubic lattice which leave 
$\pvec=\bm{0}$ invariant is the octahedral point group $O_h$, so operators
may be classified using the irreducible representations (irreps) of $O_h$.
The continuum-limit spins $J$ of our states must be deduced by examining
degeneracy patterns across the different $O_h$ irreps.

The key ingredients in our approach to extracting the hadron spectrum
of QCD from Markov-chain Monte Carlo calculations of temporal correlations
using a space-time lattice are (1) gauge field smearing using the stout-link
procedure\cite{stout}, (2) a new quark-field smearing scheme known as 
Laplacian Heaviside (\laph), (3) the use of covariantly-displaced, smeared
quark fields as our basic building blocks, (4) the use of lattice symmetry
operations and group-theoretical projections to assemble the basic building
blocks into hadron operators, and (5) the use of stochastic estimators with
diluted noise in the low-lying \laph\ subspaces to evaluate the propagators
of the basic building blocks.

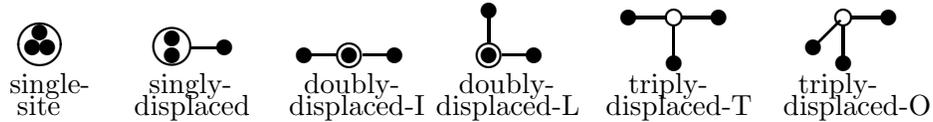
\begin{figure}[t]
\centerline{
\raisebox{0mm}{\setlength{\unitlength}{1mm}
\thicklines
\begin{picture}(16,10)
\put(8,6.5){\circle{6}}
\put(7,6){\circle*{2}}
\put(9,6){\circle*{2}}
\put(8,8){\circle*{2}}
\put(4,0){single-}
\put(5,-3){site}
\end{picture}}
\raisebox{0mm}{\setlength{\unitlength}{1mm}
\thicklines
\begin{picture}(16,10)
\put(7,6.2){\circle{5}}
\put(7,5){\circle*{2}}
\put(7,7.3){\circle*{2}}
\put(14,6){\circle*{2}}
\put(9.5,6){\line(1,0){4}}
\put(4,0){singly-}
\put(2,-3){displaced}
\end{picture}}
\raisebox{0mm}{\setlength{\unitlength}{1mm}
\thicklines
\begin{picture}(20,8)
\put(12,5){\circle{3}}
\put(12,5){\circle*{2}}
\put(6,5){\circle*{2}}
\put(18,5){\circle*{2}}
\put(6,5){\line(1,0){4.2}}
\put(18,5){\line(-1,0){4.2}}
\put(6,0){doubly-}
\put(4,-3){displaced-I}
\end{picture}}
\raisebox{0mm}{\setlength{\unitlength}{1mm}
\thicklines
\begin{picture}(20,13)
\put(8,5){\circle{3}}
\put(8,5){\circle*{2}}
\put(8,11){\circle*{2}}
\put(14,5){\circle*{2}}
\put(14,5){\line(-1,0){4.2}}
\put(8,11){\line(0,-1){4.2}}
\put(4,0){doubly-}
\put(1,-3){displaced-L}
\end{picture}}
\raisebox{0mm}{\setlength{\unitlength}{1mm}
\thicklines
\begin{picture}(20,12)
\put(10,10){\circle{2}}
\put(4,10){\circle*{2}}
\put(16,10){\circle*{2}}
\put(10,4){\circle*{2}}
\put(4,10){\line(1,0){5}}
\put(16,10){\line(-1,0){5}}
\put(10,4){\line(0,1){5}}
\put(4,0){triply-}
\put(1,-3){displaced-T}
\end{picture}}
\raisebox{0mm}{\setlength{\unitlength}{1mm}
\thicklines
\begin{picture}(20,12)
\put(10,10){\circle{2}}
\put(6,6){\circle*{2}}
\put(16,10){\circle*{2}}
\put(10,4){\circle*{2}}
\put(6,6){\line(1,1){3.6}}
\put(16,10){\line(-1,0){5}}
\put(10,4){\line(0,1){5}}
\put(4,0){triply-}
\put(2,-3){displaced-O}
\end{picture}}  }
\vspace*{8pt}
\caption{The spatial arrangements of the extended three-quark baryon
operators. Smeared quark-fields are
shown by solid circles, line segments indicate
gauge-covariant displacements, and each hollow circle indicates the location
of a Levi-Civita color coupling.  For simplicity, all displacements
have the same length in an operator.  Results presented here used
displacement lengths of $3a_s$ ($\sim 0.3 \mbox{ fm}$).
\label{fig:operators}}
\end{figure}

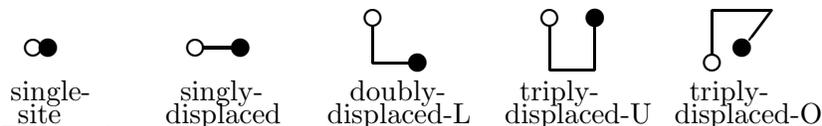
\begin{figure}[b]
\centerline{
\raisebox{0mm}{\setlength{\unitlength}{1mm}
\thicklines
\begin{picture}(20,12)
\put(7,7){\circle{2}}
\put(9,7){\circle*{2.5}}
\put(4,0){single-}
\put(5,-3){site}
\end{picture}}
\raisebox{0mm}{\setlength{\unitlength}{1mm}
\thicklines
\begin{picture}(20,12)
\put(6,7){\circle{2}}
\put(12,7){\circle*{2.5}}
\put(7,7){\line(1,0){4}}
\put(4,0){singly-}
\put(2,-3){displaced}
\end{picture}} 
\raisebox{0mm}{\setlength{\unitlength}{1mm}
\thicklines
\begin{picture}(20,12)
\put(7,11){\circle{2}}
\put(13,5){\circle*{2.5}}
\put(12,5){\line(-1,0){5}}
\put(7,10){\line(0,-1){5}}
\put(4,0){doubly-}
\put(1,-3){displaced-L}
\end{picture}}
\raisebox{0mm}{\setlength{\unitlength}{1mm}
\thicklines
\begin{picture}(20,12)
\put(8,11){\circle{2}}
\put(14,11){\circle*{2.5}}
\put(8,4){\line(1,0){6}}
\put(14,4){\line(0,1){6}}
\put(8,4){\line(0,1){6}}
\put(4,0){triply-}
\put(2,-3){displaced-U}
\end{picture}}
\raisebox{0mm}{\setlength{\unitlength}{1mm}
\thicklines
\begin{picture}(20,15)
\put(7,5){\circle{2}}
\put(11,7){\circle*{2.5}}
\put(7,12){\line(1,0){8}}
\put(7,6){\line(0,1){6}}
\put(15,12){\line(-3,-4){3.0}}
\put(4,0){triply-}
\put(2,-3){displaced-O}
\end{picture}} }
\caption[captab]{The spatial arrangements of the quark-antiquark meson operators.
In the illustrations, the smeared quarks fields are depicted by solid circles, 
each hollow circle indicates a smeared ``barred'' antiquark field, and the 
solid line segments indicate covariant displacements.
\label{fig:mesops}}
\end{figure}

We use a variety of spatially-extended hadron operators.  The use of different 
spatial configurations (see Fig.~\ref{fig:operators} for
the baryon configurations and Fig.~\ref{fig:mesops} for the meson
configurations) yield operators which effectively
build up the necessary orbital and radial structures of the hadron
excitations.  The design of these operators is such that a large number
of them can be evaluated very efficiently, and components in their
construction can be used for both meson, baryon, and multi-hadron
computations.

\begin{figure}
\begin{center}
\includegraphics[width=0.36\textwidth,clip=true]{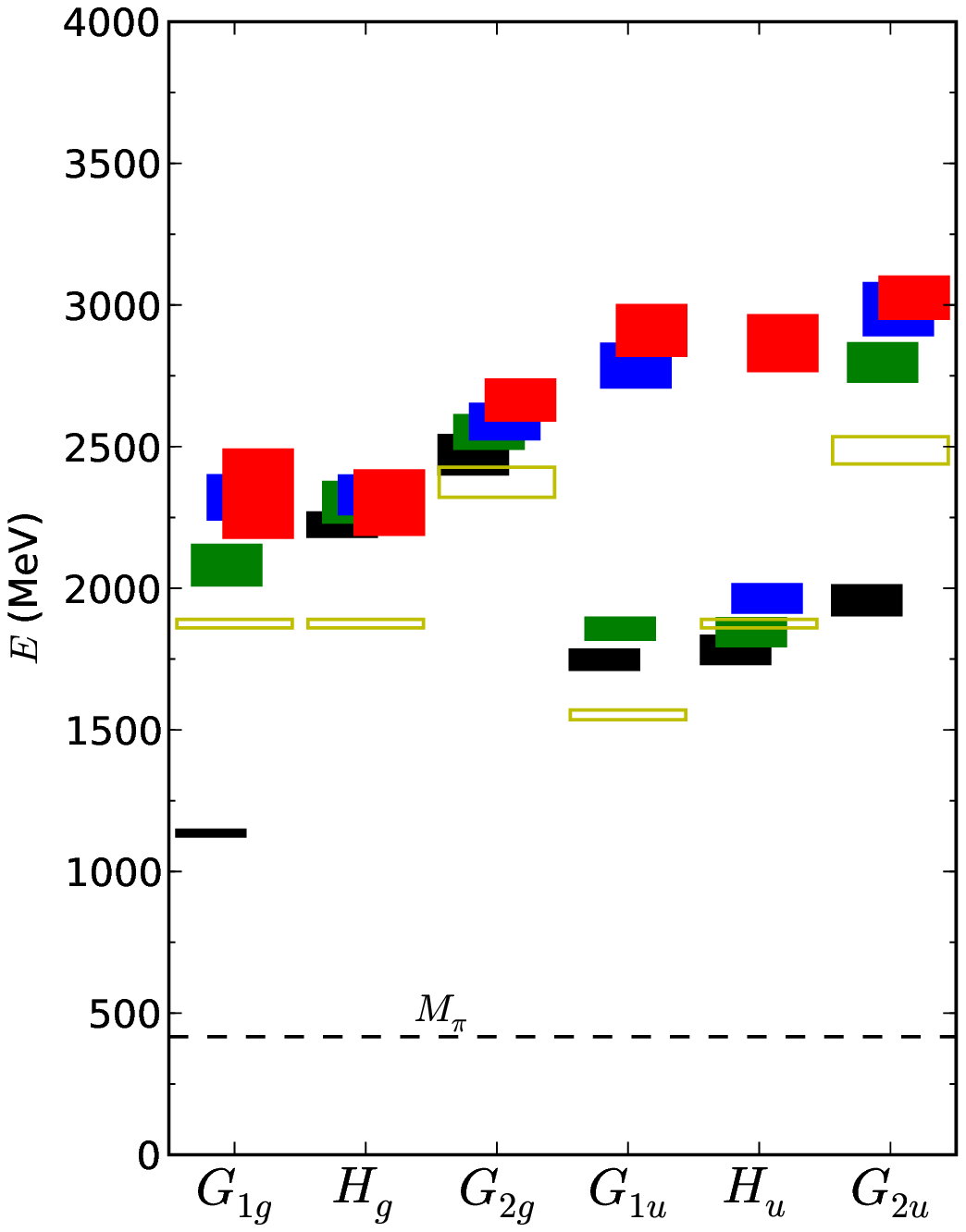}
\hspace{5mm}
\includegraphics[width=0.35\textwidth,clip=true]{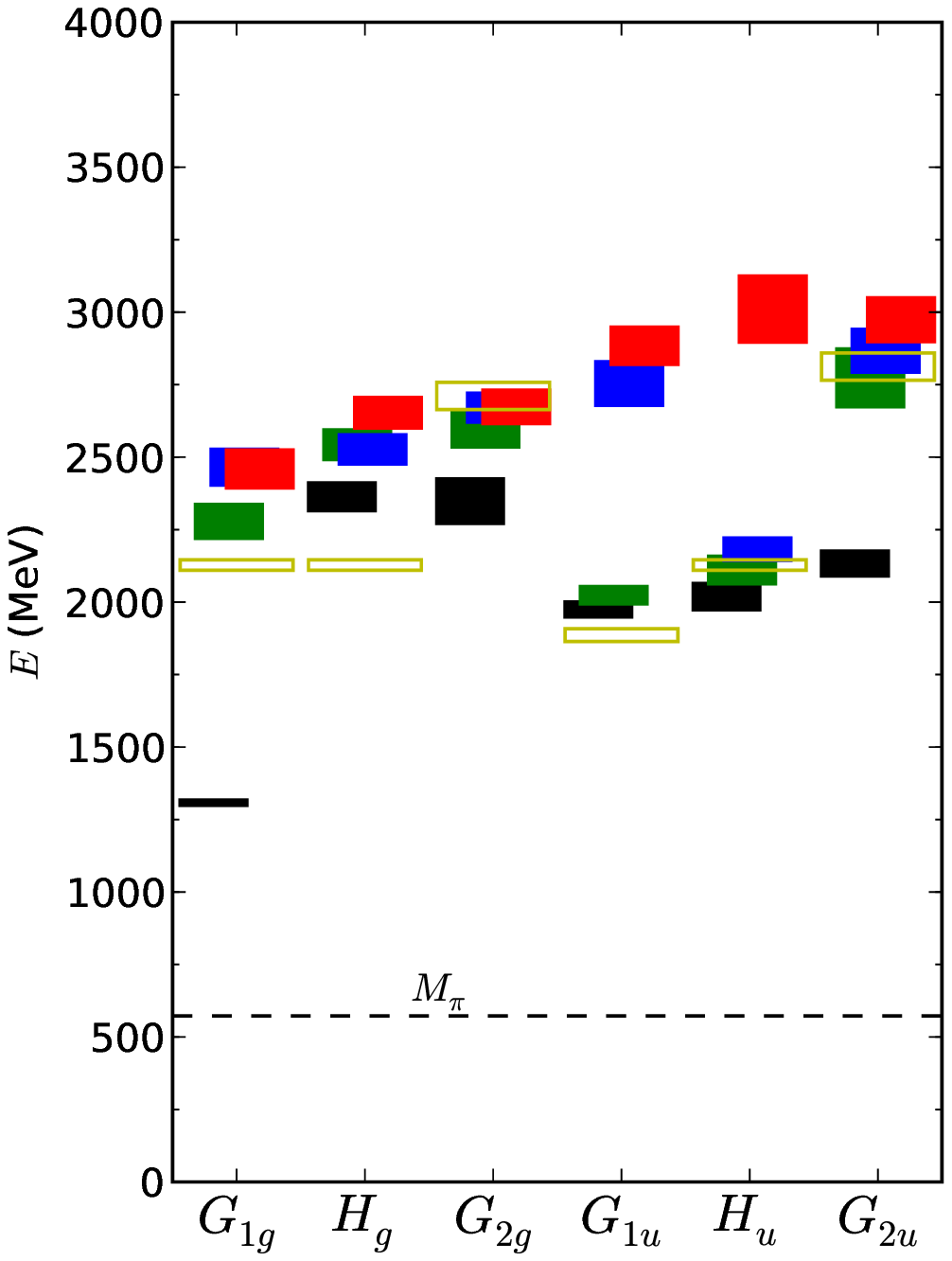}
\end{center}
\vspace*{-3mm}
\caption{The energies obtained for each symmetry channel of
isospin $\frac{1}{2}$ baryons are shown based on the
$24^3\times64$ $N_f=2$ lattice QCD data for $m_{\pi}$ = 416 
MeV (left panel) and $m_{\pi}$ = 578 MeV (right panel). 
The scale shows energies in MeV and errors are indicated by the vertical
size of the boxes. The overall error in the scale setting is not included. Empty
 boxes show thresholds for multi-hadron states.} 
\label{fig:spectrum}
\end{figure}

We recently presented results for the nucleon resonances on $N_f=2$
configurations on a $24^3\times 64$ lattice using a stout-smeared clover 
fermion action and a Symanzik-improved anisotropic gauge 
action\cite{nucleons2009} with lattice spacing $a_s\sim 0.1$~fm and 
$a_s/a_t\sim 3$.  Results, shown in Fig.~\ref{fig:spectrum}, were obtained
using 430 gauge configurations with a quark mass yielding a pion mass 
$m_{\pi}$ = 416 MeV  and using 363 gauge configurations for 
$m_{\pi}$ = 578 MeV.  The low-lying odd-parity
band shows the exact number of states in each channel as expected
from experiment.  The two figures show the splittings in the band
increasing as the quark mass is decreased.  At these heavy pion masses,
the first excited state in the $G_{1g}$ channel is significantly higher
than the experimentally measured Roper resonance.  It remains to be
seen whether or not this level will drop down as the pion mass is further
decreased.  The lowest four energies were reported in each of the six
irreducible representations of the octahedral group at each pion mass.  
Clear evidence was found for a $\frac{5}{2}^-$ state in the
pattern of negative-parity excited states.  This agrees with
the pattern of physical states and spin $\frac{5}{2}$ has been 
realized for the first time on the lattice.
Most of the levels in these plots lie very close
to two-particle thresholds, shown by empty boxes.  The use of two-hadron
operators will be needed to go to lighter pion masses.

To study a particular eigenstate of interest, all eigenstates lying below that
state must first be extracted, and as the pion gets lighter in lattice QCD 
simulations, more and more multi-hadron states will lie below the excited
resonances.  Evaluating correlators with multi-hadron sources whose 
constituents have well defined relative momenta requires quark propagators
from all lattice sites on a given time slice to all sites on another time 
slice. Computing all such elements of the propagators exactly is not possible
(except on very small lattices).  Hence, some way of stochastically estimating
them is needed.

Random noise vectors $\eta$ whose statistical expectations satisfy
$E(\eta_i)=0$ and $E(\eta_i\eta_j^\ast)=\delta_{ij}$ are useful for 
stochastically estimating the inverse of a large matrix $M$ 
as follows.  Assume that for each of $N_R$ 
noise vectors, we can solve the following
linear system of equations: $M X^{(r)}=\eta^{(r)}$ for $X^{(r)}$.
Then $X^{(r)}=M^{-1}\eta^{(r)}$, and
\begin{equation}
   E( X_i \eta_j^\ast ) = E( \sum_k M^{-1}_{ik}\eta_k \eta_j^\ast )
  = \sum_k M^{-1}_{ik}
  E(\eta_k \eta_j^\ast) = M^{-1}_{ij}.
\end{equation}
The expectation value on the left-hand can be estimated using the
Monte Carlo method.  Unfortunately, such estimates usually have
variances which are much too large to be useful.

Progress is only possible if stochastic estimates of the quark
propagators with reduced variances can be made.  Techniques of
\textit{diluting} the noise vectors have been developed which
accomplish such a variance reduction\cite{dilute1}.
A given dilution scheme introduces a complete set of $N\times N$
projection matrices $P^{(a)}$. 
Define $\eta^{[a]}_k=P^{(a)}_{kk^\prime}\eta_{k^\prime}$
and define $X^{[a]}$ as the solution of
$   M_{ik}X^{[a]}_k=\eta^{[a]}_i,$
then we have
\begin{equation}
   M_{ij}^{-1}=\sum_a M_{ik}^{-1} E(\eta^{[a]}_k \eta^{[a]\ast}_j)
      =\sum_a E(X^{[a]}_i\eta^{[a]\ast}_j).
\label{eq:diluted}
\end{equation}
The variance of such an estimate is much smaller than that without
noise dilution.  Of course, the effectiveness of the variance reduction
depends on the dilution projectors chosen.  

\begin{figure}
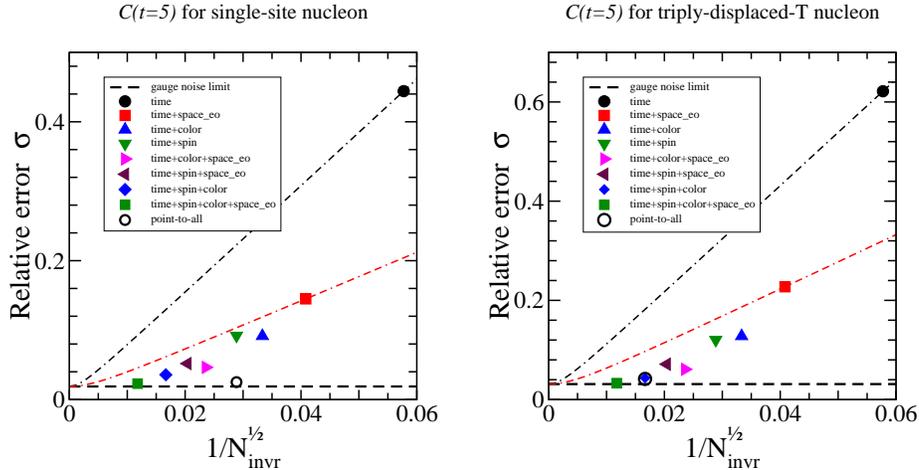

\begin{center}
\includegraphics[width=2.25in,bb=17 5 547 581]{SS_C5_dilutions.eps}
\hspace*{4mm}
\includegraphics[width=2.25in,bb=17 5 547 581]{TDT_C5_dilutions.eps}
\end{center}\vspace*{-4mm}
\caption[dilute]{
(Left) The relative errors in the correlation function of a single-site 
nucleon operator for temporal separation $t=5a_t$ evaluated using
stochastically-estimated quark propagators with different dilution
schemes against $1/N_{\rm inv}^{1/2}$, where $N_{\rm inv}$ is the
number of Dirac matrix inversions required. The open circle shows the
point-to-all error, and the horizontal dashed line shows the gauge-noise
limit.  The black (red) dashed-dotted line shows the decrease in error
expected by simply increasing the number of noise vectors, starting 
from the time (time + even/odd-space) dilution point.  (Right)
Same as the left plot, except for a triply-displaced-T nucleon operator.
These results used 100 quenched configurations on an anisotropic
$12^3\times 48$ lattice with a Wilson fermion and gauge action.
\label{fig:dilutions}}
\end{figure}

A comparison of different dilution schemes is shown in Fig.~\ref{fig:dilutions}.
The black (red) dashed-dotted line shows the decrease in error
expected by simply increasing the number of noise vectors, starting 
from the time (time + even/odd-space) dilution point.  The advantage in using
increased dilutions compared to an increased number of noise vectors with only 
time dilution is evident by the symbols lying below the dashed-dotted lines.
Note that time+spin+color+even/odd-space dilution yields an error
comparable with the gauge-noise limit using only a single noise vector!
A new smearing scheme\cite{distillation} which utilizes the low-lying
eigenvectors of the covariant Laplacian (Laplacian Heaviside smearing)
has recently been introduced,
and introducing noise vectors and dilution projectors in the subspace
spanned by the smearing produces estimates with even smaller variances.
A detailed study of this new method should be published soon.
These encouraging results demonstrate that the inclusion of good multi-hadron
operators will certainly be possible using stochastic all-to-all
quark propagators with diluted-source variance reduction.  

This talk discussed the key issues and challenges in exploring excited
hadrons in lattice QCD. The importance of multi-hadron operators
and the need for all-to-all quark propagators were emphasized.  Given the
major experimental efforts to map out the QCD resonance spectrum, 
there is a great need for \textit{ab initio} determinations of
such states in lattice QCD. The exploration of excited hadrons in lattice
QCD is well underway.

This work was supported by the National Science Foundation through awards
PHY 0653315 and PHY 0510020.

\end{document}